\documentclass[useAMS,usenatbib]{mn2e}

\usepackage{mathptmx}
\usepackage{multirow}
\usepackage{psfrag}
\usepackage{pspicture}
\usepackage{graphicx}
\usepackage{amsmath}
\usepackage{color}
\usepackage{url}
\usepackage{caption}
\usepackage{subcaption}
\usepackage{grffile}

\title[Neutral hydrogen in merging galaxy cluster]{Neutral hydrogen gas, past and future star-formation in galaxies in and around the `Sausage' merging galaxy cluster}
\author[A. Stroe et al.]{Andra Stroe$^{1}$\thanks{E-mail: astroe@strw.leidenuniv.nl}, Tom Oosterloo$^{2,3}$, Huub J. A. R\"ottgering$^{1}$, David Sobral$^{1,4,5}$\thanks{VENI/IF Fellow}, 
\newauthor Reinout van Weeren$^{6}$\thanks{Einstein Fellow}, William Dawson$^{7}$\\
$^{1}$Leiden Observatory, Leiden University, P.O.\ Box 9513, NL-2300 RA Leiden, The Netherlands\\
$^{2}$ASTRON, Postbus 2, NL-7990 AA Dwingeloo, The Netherlands\\
$^{3}$Kapteyn Astronomical Institute, Postbus 800, NL-9700 AV Groningen, The Netherlands\\
$^{4}$Instituto de Astro\'{\i}sica e Ci\^{e}ncias do Espa\c{c}o, Universidade de Lisboa, Observat\'{o}rio Astron\'{o}mico de Lisboa, Tapada da Ajuda, 1359-018, Lisbon, Portugal\\ 
$^{5}$Departamento de F\'{i}sica, Faculdade de Ci\^{e}ncias, Universidade de Lisboa, Edif\'{i}cio C8, Campo Grande, 1748-016, Lisbon, Portugal\\ 
$^{6}$Harvard Smithsonian Center for Astrophysics (CfA - SAO), 60 Garden Street Cambridge, MA 02138, US\\
$^{4}$Lawrence Livermore National Laboratory, P.O. Box 808 L-210, Livermore, CA, 94551, USA
}
\begin{document}
\maketitle
\begin{abstract}
CIZA J2242.8+5301 ($z=0.188$, nicknamed `Sausage') is an extremely massive  ($M_{200} \sim 2.0\times10^{15}M_\odot$), merging cluster with shock waves towards its outskirts, which was found to host numerous emission-line galaxies. We performed extremely deep Westerbork Synthesis Radio Telescope H{\textsc I} observations of the `Sausage' cluster to investigate the effect of the merger and the shocks on the gas reservoirs fuelling present and future star formation (SF) in cluster members. By using spectral stacking, we find that the emission-line galaxies in the `Sausage' cluster have, on average, as much H{\textsc I} gas as field galaxies (when accounting for the fact cluster galaxies are more massive than the field galaxies), contrary to previous studies. Since the cluster galaxies are more massive than the field spirals, they may have been able to retain their gas during the cluster merger. The large H{\textsc I} reservoirs are expected to be consumed within $\sim0.75-1.0$ Gyr by the vigorous SF and AGN activity and/or driven out by the outflows we observe. We find that the star-formation rate in a large fraction of H$\alpha$ emission-line cluster galaxies correlates well with the radio broad band emission, tracing supernova remnant emission. This suggests that the cluster galaxies, all located in post-shock regions, may have been undergoing sustained SFR for at least $100$ Myr. This fully supports the interpretation proposed by \citet{Stroe2015} and \citet{Sobral2015} that gas-rich cluster galaxies have been triggered to form stars by the passage of the shock. 

\end{abstract}
\begin{keywords}
galaxies: active, galaxies: clusters: individual: CIZA J2242.8+5301, shock waves, radio continuum: galaxies, radio lines: galaxies
\vspace{-15pt}
\end{keywords}

\section{Introduction}\label{sec:intro}

Galaxy cluster environments have a profound impact on the evolution of cluster galaxies. At low redshifts ($z<0.5$) and focusing on relaxed clusters, the fraction of galaxies which are star-forming drops steeply from field environments, to cluster outskirts and cores \citep{1980ApJ...236..351D, 1998ApJ...504L..75B, 2003MNRAS.346..601G}. The morphological transformation of field spirals into cluster ellipticals or S0s has been attributed to a number of processes. The dense intracluster medium (ICM) could lead to the ram pressure stripping of the gas content of field spirals as they accrete onto the cluster \citep[e.g.][]{1972ApJ...176....1G, 2014MNRAS.445.4335F}. Tidal forces produced by gradients in the cluster gravitational potential or by encounters with other galaxies, can distort infalling galaxies, truncate their halo and disk \citep[harassment,][]{1996Natur.379..613M} or remove gas contained in the galaxy and deposit it into the ICM \citep[strangulation,][]{1980ApJ...237..692L}. All these processes ultimately lead to the removal of gas and a truncation of star-formation (SF).

The effect of relaxed cluster environments on galaxies is evident using a wide range of diagnostics, which trace different phases and time-scales of SF. Using UV data produced by young OB stars, \citet{2012ApJ...750L..23O} found galaxies with star-forming trails, which they attribute to gas compression by the high-pressure merger environment. The UV radiation coming from massive, short-lived stars excites emission lines. Lines such as H$\alpha$ or [O{\textsc II}]3727{\AA} probe SF on time scales of $<10$ Myr. Emission line studies confirm that the fraction of star-forming galaxies increases from cluster cores towards field environments \citep[e.g.][]{1998AJ....115.1745G, 1998ApJ...504L..75B, 2005ApJ...630..206F,2011MNRAS.411..675S,2014ApJ...796...51D}. Using far infra-red data (tracing dust obscured SF), \citet{2012ApJ...756..106R} find that the fraction of dusty star-forming galaxies, compared to the total star-forming galaxies, increases from low to high densities, an effect they attribute to dust stripping and heating processes caused by the cluster environment. Similar results are found by \citet{2013MNRAS.434..423K}. 

Synchrotron emission from supernovae traces SF on longer timescales of about $100$ Myr \citep{1992ARA&A..30..575C}. Deep radio surveys at GHz frequencies indicate that below $100-200$ $\mu$Jy, the number of star-forming galaxies dominates over radio-loud active galactic nuclei \citep[AGN, e.g.][]{2011ApJ...740...20P}. The number of radio-faint radio star-forming galaxies \citep{2003AJ....125..506M}, in clusters was found to increase with redshift \citep{1999PhDT........22M}. Radio broad-band emission from cluster spirals was also found to be enhanced with respect to field counterparts, an effect which can be caused by compression of the magnetic fields \citep{1985ApJ...294L..89G, 1995AJ....109.1582A}. 

In addition to probes of past or current SF, CO rotational transitions can be used as an excellent tracer of molecular gas, which is the raw fuel for future SF episodes \citep[e.g.][]{2008AJ....136.2782L}. Other gas phases cannot form stars directly, but they have to cool sufficiently to form cold, dense molecular clouds \citep[see review by][]{2013ARA&A..51..105C}. However, a conversion factor between the CO mass and the total molecular gas is needed, which is highly uncertain \citep[see review by][]{2013ARA&A..51..207B}. Instead of using CO or other direct tracers of molecular gas, many studies use neutral hydrogen H{\textsc I} as a proxy for the molecular gas. Relaxed cluster spirals become increasingly more H{\textsc I} deficient towards cluster cores, an effect which does not depend on cluster global properties such as X-ray luminosity, temperature, velocity dispersion, richness or spiral fraction \citep[e.g.][]{1988ApJ...333..136M, 1990AJ....100..604C, 2001ApJ...548...97S, 2009AJ....138.1741C}. \citet{2005A&A...437L..19O} and \citet{2012MNRAS.419L..19S} have found galaxies with H{\textsc I} tails, knots and filaments, which are possibly caused by ram pressure stripping. Until very recently, H{\textsc I} measurements have been limited to the local Universe ($z\sim0$).  At low redshifts ($z\sim0.06$), \citet{2001A&A...372..768C} studied the A3128 cluster and found that the average H{\textsc I} mass for emission-line and late-type cluster members is about $(8.6-8.7)\times10^8$ M$_\odot$. They did not find a statistically significant difference between the H{\textsc I} content of emission-line galaxies inside and outside the cluster, but the field spirals contain about two times more H{\textsc I} than their cluster counterparts. Pioneering work by \citet{2007ApJ...668L...9V}, \citet{2007MNRAS.376.1357L} and \citet{2009MNRAS.399.1447L} used direct detections and stacking to measure the H{\textsc I} content of cluster galaxies up to $z\sim0.4$. \citet{2007ApJ...668L...9V} surveyed two $z\sim0.2$ clusters with very different morphologies: the relaxed, massive galaxy cluster A963 and the low-mass, diffuse cluster A2192. They detect only one H{\textsc I} galaxy within $1$ Mpc of the centre of each cluster. By stacking galaxies with known redshifts, they make a clear detection of H{\textsc I} for blue galaxies outside the clusters, but no such detection was made for cluster galaxies. In a detailed study of the cluster A370 at $z\sim0.37$, \citet{2007MNRAS.376.1357L} use spectral stacking to measure H{\textsc I} in gas-rich galaxies lying outside or at the outskirts of the cluster. 

As discussed previously, relaxed cluster environments are believed to suppress SF by removing cold gas form their host galaxies. At $z<0.3$ between $10-20$ per cent of clusters are undergoing mergers \citep{2003ApJ...585..687K, 2009MNRAS.398.1698S, 2010A&A...513A..37H} and this fraction is expected to increase steeply beyond $z=0.4$ \citep{2012MNRAS.420.2120M}. The effect of cluster mergers on the SF activity and gas content of galaxies is disputed. Most studies find that cluster mergers trigger SF \citep{Miller2003, 2004ApJ...601..805U, Ferrari2005, Owen2005, 2008MNRAS.390..289J, 2009AJ....138..873C, Hwang2009, 2014MNRAS.438.1377S, Wegner2015,Stroe2015, Sobral2015}, but a few studies find they quench it \citep[e.g.][]{2004ApJ...601..197P} or that they have no direct effect \citep[e.g.][]{2010ApJ...725.1536C}. 

An interesting subset of clusters are those hosting radio relics, extended patches of diffuse radio emission tracing merger-induced shocks \citep[e.g.][]{1998A&A...332..395E}. The H$\alpha$ properties of radio-relic clusters Abell 521, CIZA J2242.8+5301 and 1RXS J0603.3+4214 \citep{2004ApJ...601..805U, 2014MNRAS.438.1377S, Stroe2015} indicate that the merger and the passage of the shocks lead to a steep SF increase for $<0.5$ Gyr. The fast consumption of the gas ultimately quenches the galaxies within a few hundred Myr timescales \citep{2014MNRAS.443L.114R}. 

The H$\alpha$ studies of \citet{2004ApJ...601..805U}, \citet{2014MNRAS.438.1377S} and \citet{Stroe2015} are tracing instantaneous (averaged over $10$ Myr) SF and little is known about SF on longer timescales and the reservoir of gas that would enable future SF. An excellent test case for studying the gas content of galaxies within merging clusters with shocks is CIZA J2242.8+5301 \citep{2007ApJ...662..224K}. For this particular cluster unfortunately, its location in the Galactic plane, prohibits studies of the rest-frame UV or FIR tracing SF on longer timescales, as the emission is dominated by Milky Way dust. However, the rich multi-wavelength data available for the cluster give us an unprecedented detailed view on the interaction of their shock systems with the member galaxies. CIZA J2242.8+5301 is an extremely massive \citep[$M_{200}\sim 2\times10^{15} M_{\odot}$]{2015ApJ...802...46J, Dawson2015} and X-ray disturbed cluster \citep{2013PASJ...65...16A, 2013MNRAS.429.2617O, 2014MNRAS.440.3416O} which most likely resulted from a head-on collision of two, equal-mass systems \citep{2011MNRAS.418..230V, Dawson2015}. The cluster merger induced relatively strong shocks, which travelled through the ICM, accelerated particles to produce relics towards the north and south of the cluster \citep{2010Sci...330..347V, 2013A&A...555A.110S}. There is evidence for a few additional smaller shock fronts throughout the cluster volume \citep{2013A&A...555A.110S, 2014MNRAS.440.3416O}. Of particular interest is the northern relic, which earned the cluster the nickname `Sausage'. The relic, tracing a shock of Mach number $M\sim3$ \citep{2014MNRAS.445.1213S}, is detected over a spatial extent of $\sim1.5$ Mpc in length and up to $\sim150$ kpc in width and over a wide radio frequency range \citep[$150$ MHz $-16$ GHz;][]{2013A&A...555A.110S, 2014MNRAS.441L..41S}. There is evidence that the merger and the shocks shape the evolution of cluster galaxies. The radio jets are bent into a head-tail morphology aligned with the merger axis of the cluster. This is probably ram pressure caused by the relative motion of galaxies with respect to the ICM \citep{2013A&A...555A.110S}. The cluster was also found to host a high fraction of H$\alpha$ emitting galaxies \citep{2014MNRAS.438.1377S, Stroe2015}. The cluster galaxies not only exhibit increased SF and AGN activity compared to their field counterparts, but are also more massive, more metal rich and show evidence for outflows likely driven by super-novae (SN) \citep{Sobral2015}. \citet{Stroe2015} and \citet{Sobral2015} suggest that these relative massive galaxies (stellar masses of up to $\sim10^{10.0-10.7}$ $M_\odot$) retained the metal-rich gas, which was triggered to collapse into dense star-forming clouds by the passage of the shocks, travelling at speeds up to $\sim2500$ km s$^{-1}$ \citep{2014MNRAS.445.1213S}, in line with simulations by \citet{2014MNRAS.443L.114R}.

\begin{table*}
\begin{center}
\caption{Details of the JVLA observations of the `Sausage' cluster taken in L band ($1.5$ GHz), combining all four configurations of the telescope for a total of $>26$ h of observing time.}
\vspace{-5pt}
\begin{tabular}{lllll}
\hline\hline
& L-band A-array & L-band B-array & L-band C-array  & L-band D-array\\
\hline
Observation dates &   May 11, 2014  & Oct 31, 2013 &  Sep 2, 2013; Jul 3, 2013  & Feb 3, 2013; Jan 31, 2013; Jan 27, 2013 \\
Total used on source time (h)   & $\sim6.5$ & $\sim6.5$ & $\sim10$ & $\sim 3.5$\\
Integration time (s) & 1 & 3 & 5 & 5\\
\hline
\end{tabular}
\vspace{-10pt}
\label{tab:jvlaobs}
\end{center}
\end{table*}

In this paper we focus on the effect of the massive cluster merger and travelling shocks in the `Sausage' cluster on the H{\textsc I} content of the galaxies, tracing the gas that may fuel future SF. We place this in context of other phases of SF, averaged over short timescales ($\sim10$ Myr, H$\alpha$ data) and averaged over longer timescales ($\sim100$ Myr, radio broad band data) SF episodes.

The structure of the paper is as follows: in \S \ref{sec:obs-reduction} we discuss the observations and the reduction of the H{\textsc I}, optical and broad band data; in \S \ref{sec:results} we discuss the H{\textsc I} detections, stacking, masses and how these correlate with H$\alpha$ and radio luminosities; in \S \ref{sec:discussion} we discuss the implication of our results for future SF episodes and compare them with other H{\textsc I} studies of clusters. Finally, \S \ref{sec:conclusion} presents a summary of the results, placing them in context of the SF history of the cluster. At the redshift of the `Sausage' cluster, $z\sim0.188$, $1$ arcsec covers a physical scale of $3.18$~kpc and the luminosity distance is $d_\mathrm{L} \approx 940$ Mpc. All coordinates are in the J2000 coordinate system. We use a \citet{2003PASP..115..763C} initial mass function (IMF) throughout the paper. We correct measurements from other papers accordingly.

\section{Observations \& Data Reduction}
\label{sec:obs-reduction}
In our analysis, we combine radio spectral line data tracing H{\textsc I}, broad-band radio data, optical imaging and spectroscopy of passive and star-forming galaxies in and around the cluster.

\subsection{H{\textsc I} data}\label{sec:obs:HI}

The `Sausage' cluster was observed with the Westerbork Synthesis Radio Telescope\footnote{\url{http://www.astron.nl/radio-observatory/astronomers/observing-wsrt/observing-wsrt}} in the maxi-short configuration during the second half of 2012, for a total of $26$ $12$-h tracks. The observations were taken in 4, slightly-overlapping bands of $10$ MHz each and central frequencies $1180$, $1188$, $1995$ and $1203$ MHz, respectively, therefore fully covering the $1175-1208$ MHz range. Re-circulation was used to have 512 channels per band (with XX and YY polarisations only). Sources CTD93 and 3C\,147 were used as flux calibrators.

The velocity resolution (after Hanning smoothing), at the redshift of the cluster, is about $20$ km s$^{-1}$, sampled a channel width of $\sim9.9$~km~s$^{-1}$. The velocity range covered is about $4700$ km s$^{-1}$, corresponding to an H{\textsc I} redshift range of $0.184-0.199$. The redshift range covers the cluster volume within $-1\sigma$ to $+3\sigma$ of the cluster redshift $z=0.188$, where $\sigma=0.04$ is the cluster velocity dispersion \citep{Dawson2015}. The H{\textsc I} observations cover well the distribution of the H$\alpha$ emitting galaxies \citep[$z=0.190\pm0.010$;][]{Sobral2015}.

Significant radio frequency interference (RFI) is known to be present at the WSRT at frequencies covering the H{\textsc I} redshift range of $0.1$ to $0.25$, caused by geo-positional systems such as GPS and GLONASS. However, at the time of our observations, the frequencies corresponding to the redshift of the cluster ($1185-1200$ MHz) were still fairly free of such RFI. With the foreseen deployment of the GALILEO geo-positional system, the situation in this frequency range will worsen dramatically in the near future.

To remove any residual RFI, we performed our RFI flagging on the Stokes Q (i.e.\ XX-YY) component on the data. Given the polarised nature of RFI, this removed most of the astronomical signal, but left the RFI mostly intact. The flags found for Stokes Q were then applied to original XX and YY visibilities. Moreover, because the RFI in our data is broad band, before flagging we performed a smoothing in frequency to enhance the `sensitivity' for RFI. These procedures worked very well and the final data cubes do not show any effects of residual RFI, while the noise level ($75$ $\mu$Jy beam$^{-1}$ over $20$ km s$^{-1}$) is very close to what is expected for the integration time. 

Once the data were flagged, we applied standard calibration procedures to the data using the {\textsc miriad} software \citep{Sault1995}. The continuum emission was removed from the data by fitting, to all channels, a 3$^{\rm rd}$ order polynomial to each visibility spectrum (`uvlin'). The choice of the order depends on out to which radius there are significant continuum sources and on bandwidth. The higher the order, the better the sources are removed, however the noise, after subtraction, increases. A 3$^{\rm rd}$ order polynomial fit represents a compromise between sufficient removal of the continuum and little increase in the noise level of the line cube, as shown by \citet{1994A&AS..107...55S}. Because this does not take into account the presence of any possible H{\textsc I} emission, the spectra of individual detections, and of the stacked spectra we discuss below, are corrected for this over-subtraction (see Section~\ref{sec:results:cont}).

The synthesised beam of the WSRT observations is $24.9\times18.5$ arcsec$^2$ at a position angle of $165.8^{\circ}$, or $79.3\times 58.9$ kpc$^2$.

\begin{table*}
\begin{center}
\caption{Number of sources with spectroscopic data, separated by galaxy type. Not all spectroscopic sources are covered by the redshift range of the H{\textsc I} observations.}
\vspace{-5pt}
\begin{tabular}{l c c c}
\hline
\hline
Sample & Total number & Sources within H{\textsc I} $z$ range & Reference \\  \hline
Emission line, field & \phantom{$0$}$39$  & \phantom{$0$}$22$ & \citet{Sobral2015,Stroe2015}  \\
Emission line, cluster & \phantom{$0$}$54$ & \phantom{$0$}$45$ & \citet{Sobral2015,Stroe2015} \\
Cluster star-forming & \phantom{$0$}$48$ & \phantom{$0$}$39$ & \citet{Sobral2015} \\
Cluster AGN & \phantom{$00$}$6$ & \phantom{$00$}$6$ & \citet{Sobral2015} \\
All emission-line  & \phantom{$0$}$93$ & \phantom{$0$}$67$ & \citet{Sobral2015,Stroe2015} \\
Passive, cluster & $184$ & $154$ & \citet{Dawson2015} \\ \hline
All & $277$ & $221$  & \\
\hline
\end{tabular}
\vspace{-10pt}
\label{tab:spectra}
\end{center}
\end{table*}

\subsection{Optical imaging and spectroscopy}\label{sec:obs:optical}

In order to locate the spatial and velocity position of the H{\textsc I} signal, we use the Keck and William Herschel Telescope spectroscopy. Data of both passive and star-forming galaxies in the field of the cluster were presented in \citet{Sobral2015}, \citet{Dawson2015} and \citet{Stroe2015}. 

\begin{figure}
\begin{center}
\includegraphics[trim=0cm 0cm 0cm 0cm, width=0.495\textwidth]{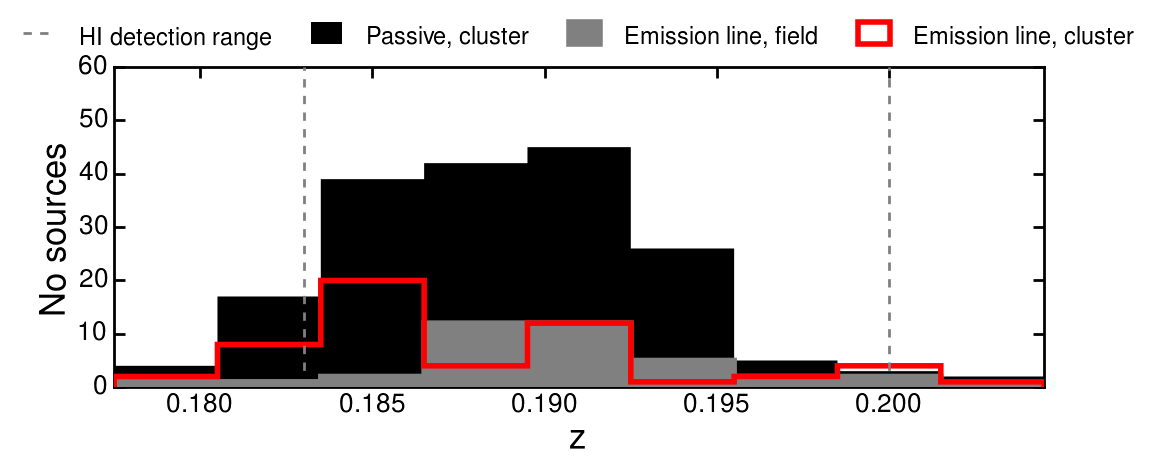}
\end{center}
\vspace{-10pt}
\caption{Redshift distribution of sources in each galaxy sample (emission-line cluster members, emission-line field galaxies and cluster passive galaxies). The H{\textsc I} redshift coverage of the WSRT data is marked by the vertical dashed lines.}
\vspace{-10pt}
\label{fig:histz}
\end{figure}

Galaxies are categorised as passive or emission-line based on spectral features. Emission line galaxies were selected through the presence of the H$\alpha$ emission-line (with a H$\alpha$+[N{\textsc II}] rest-frame equivalent width larger than $13${\AA}), tracing hot ionised gas, indicating the presence of SF and/or radio-quiet AGN (broad and narrow line AGN). The spectra of passive galaxies display Balmer absorption features and/or no Balmer series emission lines. The passive galaxies are undetected in H$\alpha$ at the $13$ {\AA} level equivalent width level. The sample was divided into three categories:
\begin{enumerate}
\item passive galaxies inside the cluster,
\item emission-line galaxies within the cluster,
\item emission-line galaxies in the field around the cluster. 
\end{enumerate}
The number of sources in each sub-sample can be found in Table~\ref{tab:spectra}. The emission-line galaxy sample is dominated by star-forming galaxies with a $\sim20-30$ per cent contribution from broad and narrow line optical AGN \citep[see Table~\ref{tab:spectra};][]{Sobral2015}. The cluster members were chosen to be located at a projected distance of less than $1.85$ Mpc away from the cluster `centre', in line with the definition of cluster membership from \citet{Stroe2015}. Line-emission galaxies outside the cluster were defined to lie outside of the $1.85$ Mpc radius. The redshift distribution of the galaxies is plotted in Figure~\ref{fig:histz}. We note that, as shown by \citet{Sobral2015}, the samples of cluster and field line emitters are selected uniformly, down to a similar star-formation rate (SFR).

The spectroscopy is supplemented with Subaru, Canada-France-Hawaii, William Herschel Telescope and Isaac Newton Telescope broad band (BB) and narrow-band (NB) imaging tracing the H$\alpha$ emission-line in galaxies at the cluster redshift \citep{Stroe2015}. H$\alpha$ luminosities for each source were calculated using the excess of NB emission compared to the $i$ BB filter \citep[for method and details see][]{Stroe2015}. In the analysis, we also employ stellar masses derived using the method described in \citet{Sobral2015}.

\subsection{Broad-band radio data}
We identified radio counterparts to the optical sources by cross-matching with a deep, high resolution ($\sim1.5$ arcsec) image of the cluster, centred at $1.5$ GHz, produced using the upgraded Jansky Very Large Array.

Deep JVLA observations of the cluster were taken in the $1-2$ GHz L-band in A, B, C, and D-array configurations. An overview of the observations is given the Table~\ref{tab:jvlaobs}. In total, $16$ spectral windows with $64$ channels, each covering $64$~MHz of bandwidth, were recorded. The data reduction for each observing run was carried out separately, using {\tt CASA}\footnote{http://casa.nrao.edu} version 4.2.

As a first step, the data was Hanning-smoothed and corrections for the antenna positions and elevation dependent gains were applied. We then obtained an approximate bandpass solution using observations on the primary calibrators (3C147, 3C138). We applied the bandpass solutions to the data and flagged RFI in an automated way using the {\tt AOFlagger} \citep{2010MNRAS.405..155O}. This initial bandpass correction was performed to avoid flagging of data due to the bandpass variations across the spectral windows. After flagging, we obtained gain corrections on the primary calibrators using $5$ channels centred at channel $30$. These gain solutions were obtained to remove the time-varying gains. We pre-applied these solutions to find delay terms and bandpass solutions. We then re-determined the gain solutions but now using the full channel range pre-applying the bandpass and delay corrections. We then pre-applied these solutions to obtain gain solutions on our secondary calibrator J0542+4951. The cross-hand delays were solved for using the calibrator 3C\,138. The channel dependent polarization leakage terms and polarization angles were set using 3C\,147 and 3C\,138, respectively. For observing runs longer than four hours the leakage terms were determined from scans on the secondary calibrator J0542+4951\footnote{The polarization results will be discussed in a forthcoming paper (van Weeren et al. in prep.)}. 

We bootstrapped the flux-scale from our primary calibrator observations to find the flux-density of J0542+4951. The flux-scale was set using the default settings of the task {\tt setjy}. As a final step, the calibration tables were applied to the target data. We averaged the target field data by a factor of $3$ in time and frequency, to reduce the data volume for imaging.

In the next steps, the calibration solutions were refined using several cycles of phase-only and amplitude and phase self-calibration. For the imaging we employed w-projection  \citep{2008ISTSP...2..647C,2005ASPC..347...86C} and MS-MFS clean \citep{2011A&A...532A..71R} with {\tt nterms=3}. Clean boxes were set by running the Python Blob Detection and Source Measurement ({\tt PyBDSM}\footnote{http://dl.dropboxusercontent.com/u/1948170/html/index.html}). A few additional clean regions were added manually for diffuse sources.

After self-calibrating the individual datasets, we combined all the datasets to make one deep image. Two more rounds of amplitude and phase self-calibration (on a $1$ h timescale) were carried out on the combined dataset. During the self-calibration the amplitude scale was allowed to drift freely. This was needed to fully align the different datasets and spectral windows and avoid strong artefacts around a few bright sources located in the field of view (FOV)\footnote{In principle, the global amplitude scale could have been preserved, but such an option is not offered in {\tt CASA} at the moment.}. We made a final image using Briggs ({\tt robust=0}) weighting and corrected for the primary beam attenuation. We checked the flux-scale of the image against our previous $1.38$~GHz WSRT observations of the cluster \citep{2010Sci...330..347V, 2013A&A...555A.110S}. This was done by checking the integrated fluxes of $10$ sources and scaling the fluxes from $1.4$ to $1.5$ GHz assuming a spectral index of $-0.7$,  the canonical spectral index for bright radio sources \citep[e.g.][]{1992ARA&A..30..575C}. The JVLA fluxes were divided by a correction factor of $1.4\pm0.1$ to re-align the flux-scale to the WSRT scales. The resolution of the VLA image is $1.5$ arcsec $\times1.4$ arcsec, with a position angle of $86.5$ degrees.

\section{Results}\label{sec:results}

\begin{table} 
\caption{Properties of the direct detections. The positional uncertainties are $15$ arcsec, the error in the redshifts is 0.0005 and the error in the H{\textsc I} mass about $0.4 \cdot 10^9$ M$_\odot$. The stellar masses reported are for the closest face-on spiral found within 7 arcsec.}
\vspace{-5pt}
\centering 
\begin{tabular}{cccccc} 
\hline\hline
RA & DEC                       & $z$   & $M_{\rm HI}$   & $M_\star$  & $M_{\rm HI}/M_\star$ \\
  (h m s)        & ($^\circ$ $^\prime$ $^{\prime\prime}$)  &       & ($10^9$ M$_\odot$)    & ($10^9$ M$_\odot$)  &   \\
\hline
22 41 51.1 & 52 52 55  & 0.18893   & $2.1$ & $2.0$ & $1.0$ \\
22 42 56.5 & 52 57 21  & 0.18916   & $2.6$ & $^*$  & -- \\ 
22 43 14.3 & 53 04 57  & 0.18536   & $1.2$ & $9.7$ & $0.1$ \\
22 43 23.2 & 53 04 41  & 0.18536   & $2.2$ & $^{**}$ & -- \\
22 41 30.4 & 53 05 58  & 0.18486   & $1.7$ & $^{**}$ & -- \\
22 43 43.7 & 53 09 43  & 0.18994   & $1.6$ & $7.9$ & $0.2$ \\
\hline
\end{tabular}\\
{\small $^*$ A bright star overlaps the position of the galaxy, so no reliable counterpart can be found.\\
 $^*$ No face-on spiral was found nearby the H{\textsc I} detection indicating the detection is spurious.\\}
\vspace{-10pt}
\label{tab:radio}
\end{table}

\subsection{Direct detections}
Six galaxies are tentatively detected in H{\textsc I}. The detection criterion is signal above $5\sigma$ over at least two consecutive channels. Table \ref{tab:radio} lists the redshifts and H{\textsc I} masses of the direct detections. The narrow H{\textsc I} profiles ($\sim40$ km s$^{-1}$) of the directly detected sources indicates, if real, they are most probably oriented in the plane of the sky. For a given H{\textsc I} mass, the peak flux for a face on galaxy is higher than for an edge-on galaxy because the same amount of flux divided up in fewer channels. Therefore, we are biased towards detecting face-on galaxies. However, given the very narrow profiles of these tentative detections, they could also be high noise peaks.

We note that none of the six H{\textsc I} direct detections have counterparts (matches within $10$ arcsec) in our spectroscopic data or in the H$\alpha$ catalogue. This indicates that the sources have faint H$\alpha$ emission, below the equivalent width detection threshold ($13${\AA}). Their SFR are therefore below $\sim0.35$ $M_\odot$ yr$^{-1}$. 

Given the positional accuracy coupled with the large beam of WSRT finding the right optical counterpart is very challenging. A few ($1-6$) potential optical hosts are found for the tentative H{\textsc I} direct detections, but most sources are unresolved and faint ($i$ band magnitude on average fainter than $20$). Therefore reliable photometric redshifts could not be derived and we cannot confirm them as being located at $z\sim0.2$. However, even if these sources were $z\sim0.2$ galaxies, they would have small stellar masses. For example, using the closest optical match as galaxy host, we find that their stellar masses are very small ($<1 \times 10^9 M_\odot$). This would imply unrealistically high gas to stellar mass ratios. Additionally, the morphology of these close optical sources does not match face-on spiral galaxies, as we expect. These arguments indicate that the closest sources are not the correct matches. For three out of the six H\textsc{I} detections there are large, face-on spiral galaxies in their vicinity (within $7$ arcsec), which could be the correct optical counterpart. These three galaxies have stellar masses of $\sim(2-10) \times 10^9$ $M_\odot$, which indicates $0.1-1$ atomic gas to stellar fractions. For two detections only small sources are located in the vicinity and no obvious face-on galaxies are found nearby the H{\textsc I}, indicating these are spurious detections (noise peaks in adjacent channels mimicking a signal). In one case, a bright star located at the location of the H{\textsc I} detection prevent correct optical identification.

\subsection{H{\textsc I} stacking}\label{sec:results:stacking}
Since only six galaxies were directly detected, we use spectral stacking to measure the average H{\textsc I} content of the galaxy samples. 

We use the optical positions to extract radio spectra for each galaxy, summing the flux within an elliptical aperture equal to the FWHM of the synthesised beam ($24.9\times18.5$ arcsec$^2$). This corresponds to a spatial scale of $79.3\times58.8$ kpc$^2$, matched to the physical size of the galaxies, which at $z\sim0.19$ are unresolved in the H{\textsc I} observations \citep{2007ApJ...668L...9V} (also in line with the galaxy sizes presented in \citealt{2014MNRAS.438.1377S, Stroe2015}). We further tested the effect of the aperture size on the final stack using apertures of sizes ranging from $0.1$ of the FWHM up to $2$ times the FWHM.

\begin{figure}
\begin{center}
\includegraphics[trim=0cm 0cm 0cm 0cm, width=0.495\textwidth]{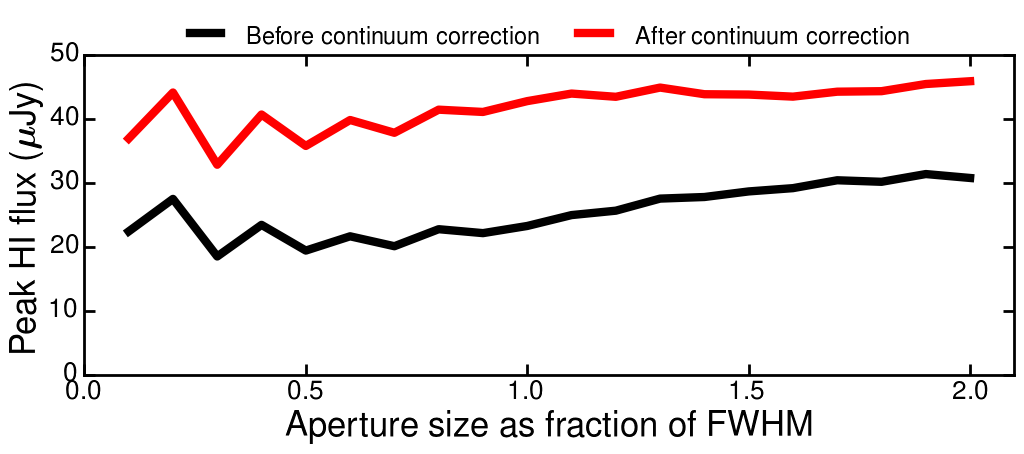}\\
\includegraphics[trim=0cm 0cm 0cm 0cm, width=0.495\textwidth]{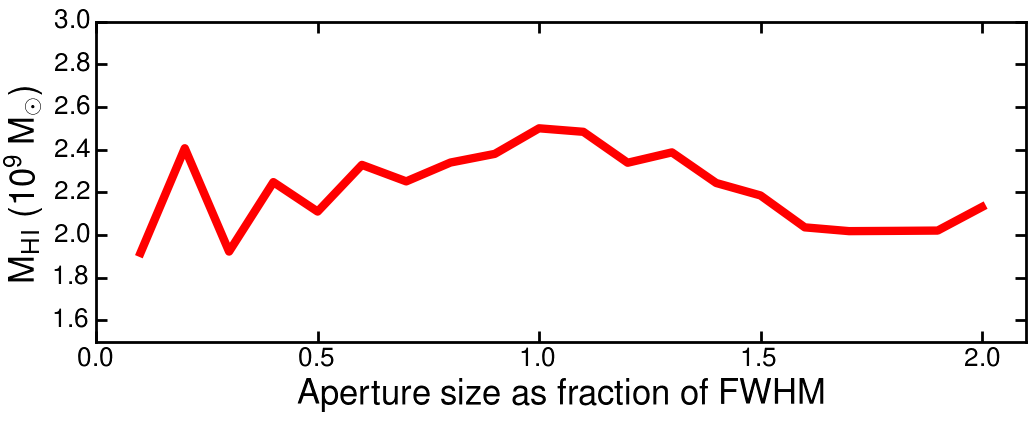}
\end{center}
\vspace{-10pt}
\caption{Top: H{\textsc I} peak emission flux in the cluster line emitter stack as function of the aperture size used for extracting the individual spectra. The horizontal scale indicates the fraction by which the aperture width and height are scaled as a function of the FWHM. The trend in the peak emission does not change before and after correcting for the continuum over-subtraction (see \S\ref{sec:results:cont}). Bottom: Average H{\textsc I} mass of cluster line emitters (see \S\ref{sec:results:HI}), as function of the aperture size used for extracting the individual spectra. The horizontal scale indicates the fraction by which the aperture width and height are scaled as a function of the FWHM.}
\vspace{-10pt}
\label{fig:curve}
\end{figure}

To test the effect of aperture size on the final H{\textsc I} stack, we use elliptical apertures in size equal to a fraction of the FWHM on both the width and height of the synthesised beam. We use apertures from 0.1 to 2 times the FWHM of the radio beam. We follow the procedure described in \S~\ref{sec:results:stacking} to extract spectra at the positions of the cluster emission-line members. We find that the peak of the H{\textsc I} detection remains relatively stable if the aperture is at least $0.9$ of the FWHM size (see top panel, Figure~\ref{fig:curve}).

For each aperture size (from 0.1 to 2.0 times the FWHM of the radio beam, see Figure \ref{fig:curve}), we also measure the H{\textsc I} mass in the way described in \S \ref{sec:results:HI}. As shown in the bottom panel of Figure~\ref{fig:curve}, we find that the H{\textsc I} mass is relatively stable as function of aperture, but it peaks when the size of the aperture is equal to the FWHM of the synthesised beam. The FWHM size of WSRT is also well matched to the expected H{\textsc I} disk size of galaxies at $z\sim0.19$ \citep{2007ApJ...668L...9V}. We therefore choose to measure H{\textsc I} quantities within apertures equal to the FWHM of the synthesised beam.

We extract spectra for all galaxies whose redshift falls within the H{\textsc I} redshift range probed by our WSRT observations ($0.184<z<0.199$). To study the noise properties, we extract spectra in sky positions shifted by $-60$ arcsec in the RA direction, but using the same redshifts as the sample of galaxies. This method captures the effect of increasing noise towards the edges of the bandpass. Note however that the noise could be overestimated. Given the large source density of the cluster field, a shift in sky position does not guarantee we will be measuring pure noise, but some apertures could partially fall on undetected source. 

Before stacking, we correct the galaxy and noise spectra for the WSRT primary beam, which is a function of distance from the pointing centre and observing frequency:
\begin{equation}
A(r, \nu)= \cos^6(c \nu r)
\end{equation}
where $c = 68$ is a constant, $r$ is the distance from the pointing center in degrees and $\nu$ is the observing frequency in GHz. By correcting for the primary beam, we account for the effect of noise increasing towards the FOV edges in both the galaxy and their associated noise spectra.

The spectra are then shifted to the rest-frame velocity using their spectroscopic redshift. We use the radio definition of velocity: $V = c (\nu_\mathrm{HI}-\nu_\mathrm{observed})/\nu_\mathrm{HI}$, where $c$ is the speed of light, $\nu_\mathrm{HI}$ is the restframe frequency of H{\textsc I} and $\nu_\mathrm{observed}$ is the observing frequency. The galaxy spectra and the noise spectra are co-added using a weight based on the primary beam ($\propto A^2(r, \nu)$), which accounts for the fact that the noise for galaxies away from the field centre is larger in proportion to the primary beam. To obtain the correct flux density scale, we normalise the stacked spectrum by the integral of the synthesised beam, integrated over the same spatial region used for extracting the galaxy spectra.

\begin{figure}
\begin{center}
\includegraphics[trim=0cm 0cm 0cm 0cm, width=0.495\textwidth]{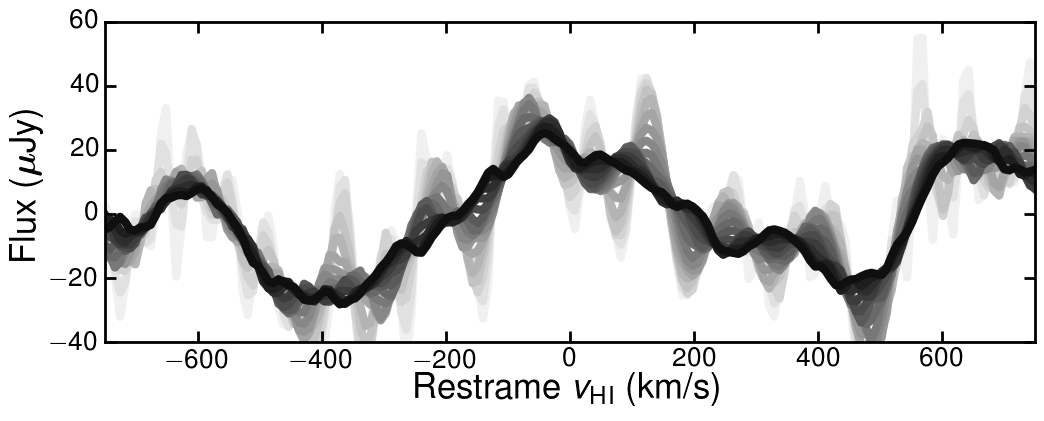}
\end{center}
\vspace{-10pt}
\caption{Noise reducing properties of the Savitzky-Golay filter with increasing filter window width demonstrated on the cluster member galaxy stack. Colours of the lines go from lighter to darker shades in increasing window length. The black line at the foreground of the figure is smoother spectrum after applying the final window length choice, $\sim300$ km s$^{-1}$, which is also the one we finally applied to the data. Similar results are obtained for the galaxy stacks, as well as their associated noise stacks.}
\vspace{-10pt}
\label{fig:sg}
\end{figure}

After stacking and normalising the spectra, we filter the data using a second-order Savitzky-Golay  filter \citep[SG;][]{1964AnaCh..36.1627S}, which convolves the data with a polynomial filter. Given that our line profiles are resolved (FWHM of the stacks with detection are $>150$ km s$^{-1}$, compared to a channel width of $>20$ km s$^{-1}$, see Table \ref{tab:stacked}), the method reduces the noise, while preserving line profiles \citep[see for example][]{2014ApJ...795L..33M}. We tested the method using different window sizes. The effect of the SG filtering with increasing window lengths is shown in Figure~\ref{fig:sg}. Out of the window size tested, we finally choose a filter window of $\sim300$ km s$^{-1}$, which provides minimal noise, while preserving the width and height of the signal, thus maximising the signal to noise of the possible H{\textsc I} detection peaks. We also tested other smoothing kernels (e.g. moving boxcar), which resulted is similar results, but with a widening of the profile and reduction of the peak strength.

Separate stacks are produced for the sample of passive cluster galaxies, line emission cluster members and line emission galaxies located within the field environment around the cluster. Line emitters include both star-forming galaxies and optical AGN (see also \S~\ref{sec:obs:optical}). Finally, we produce a master stack of all the galaxies available. The number of galaxies per each velocity channel for each stack is shown in Figure~\ref{fig:nostacked} and Table~\ref{tab:stacked}. The number of sources in the stack naturally peaks at the $0$-velocity position, but dwindles towards higher relative velocities. This effect is governed by where the redshift of each source falls within the WSRT HI bandpass. Due to extensive spectroscopy from Keck aimed at obtaining a dynamical analysis of the cluster that specifically targeted the red sequence, the number of passive cluster galaxies far outnumbers the number of emission-line galaxies. 

The asymmetry (about the central position) in the number of sources for which data exists, especially visible in the passive cluster galaxy sample, is caused by the discrepancy in the nominal redshift of the cluster \citep[$z=0.188$, recently derived from more than 200 spectra by][]{Dawson2015} and the outdated $z=0.192$ \citep{2007ApJ...662..224K} which was used for creating the WSRT H{\textsc I} setup. Our spectroscopic measurements are therefore biased towards lower redshifts, compared to the HI coverage of the WSRT data (see \S\ref{sec:obs:optical} and Figure \ref{fig:histz}). If a redshift of a source falls at the middle of the WSRT H{\textsc I} band coverage, the frequency coverage is symmetric about the observed frequency of the H{\textsc I}. However, a lower redshift (than the central redshift) is equivalent to a source having a wider frequency coverage at frequencies lower than the H{\textsc I} and a narrower coverage at higher frequencies. When translating to a rest-frame velocity, there is a preferential data coverage of the positive restframe velocities. The missing data at larger absolute restframe velocities drives the noise to higher values in those regions. 

\subsection{Correcting for the over-subtracted continuum emission}\label{sec:results:cont}
As mentioned previously, the bulk of our galaxies do not have a direct detection of the H{\textsc I} line. Hence, the continuum emission subtraction (see \S \ref{sec:obs:HI}) could not take into account the presence of H{\textsc I} emission and leads to an over-subtraction of the continuum where the putative H{\textsc I} line is located. The over-subtraction is not visible (and relevant) in the individual spectra, but it shows up in the stacked spectrum. As expected, the over-correction of the continuum is evident in the case of the emission-line galaxy samples, where the H{\textsc I} signal is located on top of a broad, negative dip. No evident negative trough is present in the passive member sample, where less H{\textsc I} is expected.

\begin{figure}
\begin{center}
\includegraphics[trim=0cm 0cm 0cm 0cm, width=0.495\textwidth]{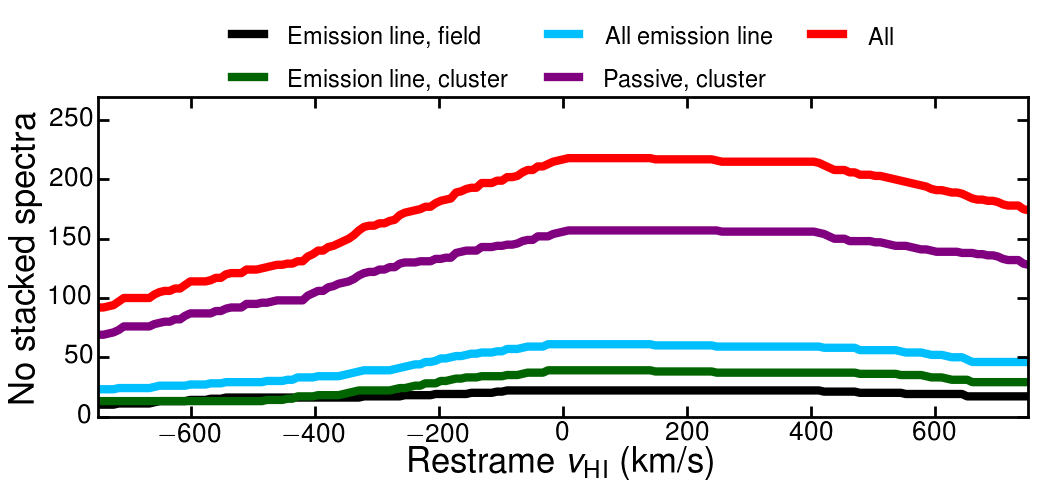}
\end{center}
\vspace{-10pt}
\caption{Number of sources stacked in each velocity channel, for the separate stacks. The passive cluster galaxy stack (purple line) benefits from a factor of $>5$ more sources than the stacks for the H$\alpha$ line emitters (mostly star-forming and some AGN) galaxies inside (green line) and outside the cluster (black line). Note how the number of sources dwindles towards the edges of the WSRT H{\textsc I} bandpass. The asymmetry in the number of sources is caused by the redshift distribution of the stacked sources (see Figure \ref{fig:histz}), preferentially located at redshifts lower than the central WSRT HI coverage.}
\vspace{-10pt}
\label{fig:nostacked}
\end{figure}

We apply a two step process to correct for the over-subtraction of the continuum. Firstly, to measure the possible extent of the H{\textsc I} in the H$\alpha$ galaxies, we fit Gaussian profiles to their stacked profiles. We select data at least $4\sigma_\mathrm{Gauss}$ away from the peak of the Gaussian, where $\sigma_\mathrm{Gauss}$ is the dispersion of Gaussian profile. Additionally, we discard the data at relative restframe velocities higher than $1000$ km s$^-{1}$. These cuts are employed to exclude any H{\textsc I} signal from the estimation of the continuum, but not include very noise edge channels, where only a few galaxies are stacked (see Figure~\ref{fig:nostacked}). In the case of the passive members, a similar procedure is applied, but since we do not have a clear detection of emission, the velocity range between $250$ and $1000$ km s$^{-1}$ is used.

\begin{figure}
\begin{center}
\includegraphics[trim=0cm 0cm 0cm 0cm, width=0.495\textwidth]{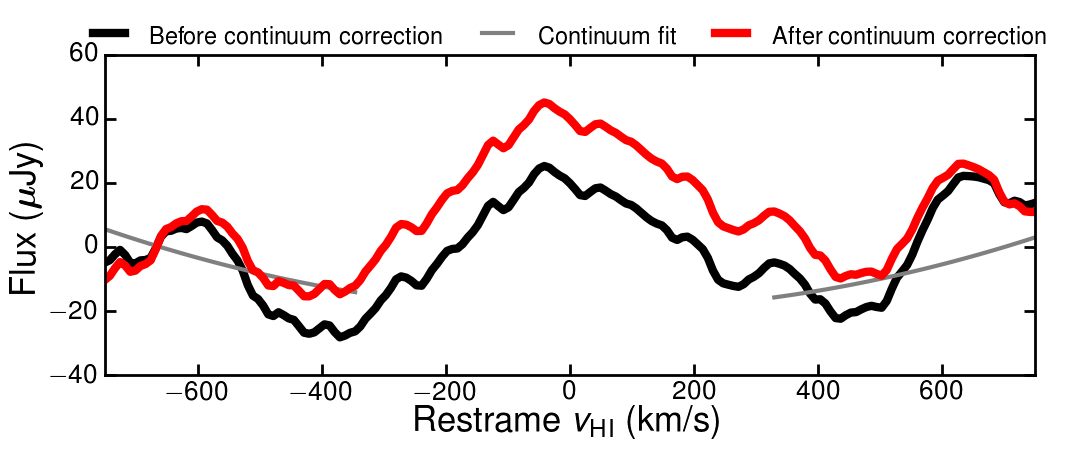}
\end{center}
\vspace{-10pt}
\caption{Correction for the continuum over-subtraction in the case of the cluster line emitter H{\textsc I} stack.}
\vspace{-10pt}
\label{fig:beforeafter}
\end{figure}

We fit the channels free of H{\textsc I} with a 2$\mathrm{nd}$ order polynomial and subtract the fit from the data to correct for the over-subtracted continuum (see example in Figure~\ref{fig:beforeafter}). The correction is substantial for the cluster H$\alpha$ line galaxies, but the results indicate that the continuum has not been significantly over-subtracted for passive cluster members and those line emitters located in the field environment.

\subsection{Measuring the H{\textsc I} signal and its significance}\label{sec:results:HI}
\begin{figure*}
\centering
\begin{subfigure}[b]{0.795\textwidth}
\includegraphics[trim=0cm 0cm 0cm 0cm, width=0.995\textwidth ]{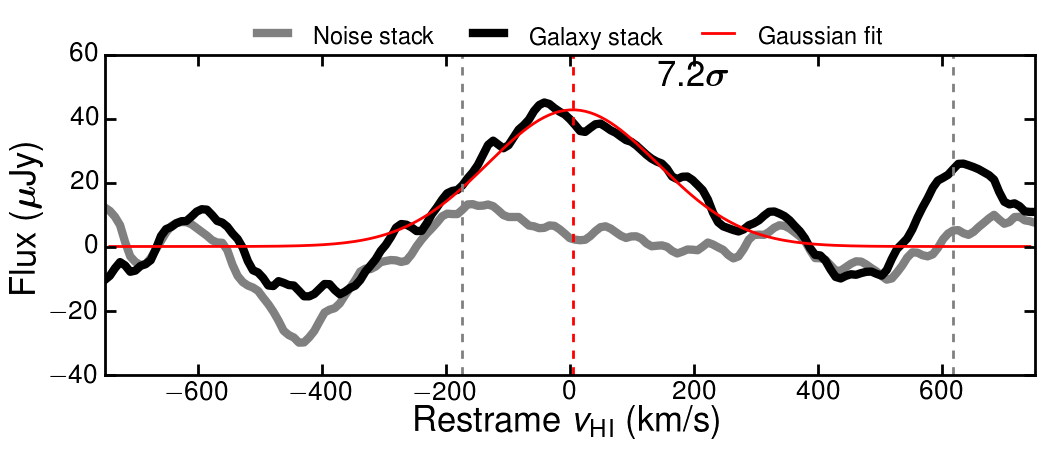}
\vspace{-10pt}
\caption{Emission line cluster galaxies.}
\label{fig:stacks:members}
\end{subfigure}
\begin{subfigure}[b]{0.495\textwidth}
\includegraphics[trim=0cm 0cm 0cm 0cm, width=0.995\textwidth ]{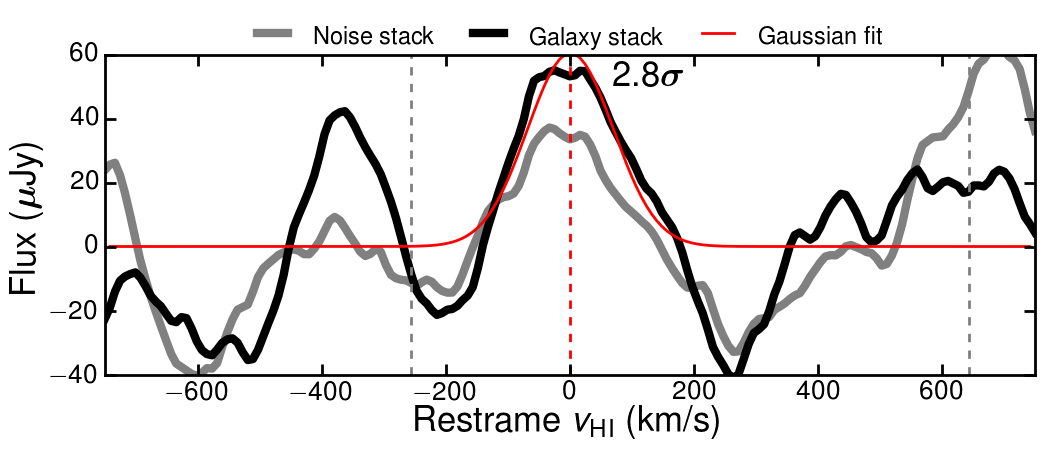}
\vspace{-10pt}
\caption{Emission line field galaxies.}
\label{fig:stacks:nonmembers}
\end{subfigure}
\begin{subfigure}[b]{0.495\textwidth}
\includegraphics[trim=0cm 0cm 0cm 0cm, width=0.995\textwidth ]{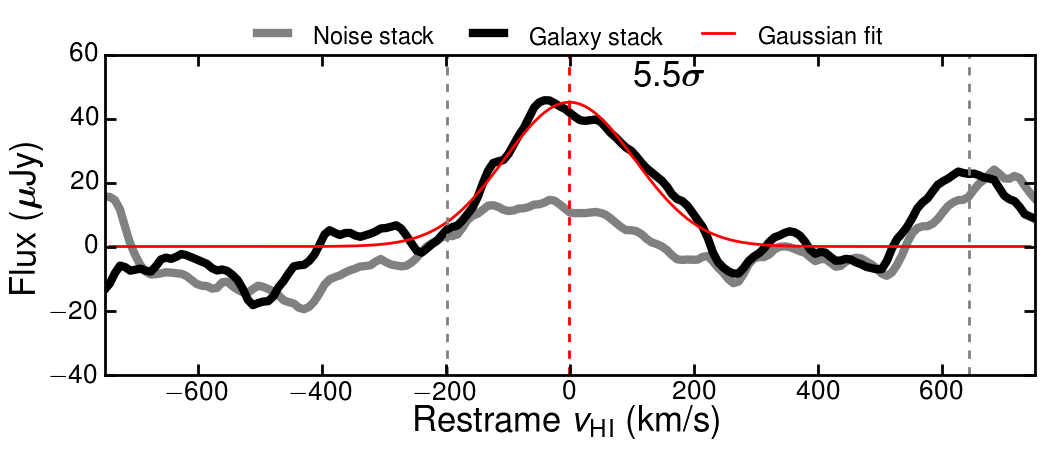}
\vspace{-10pt}
\caption{All emission-line galaxies.}
\label{fig:stacks:sf}
\end{subfigure}
\begin{subfigure}[b]{0.495\textwidth}
\includegraphics[trim=0cm 0cm 0cm 0cm, width=0.995\textwidth ]{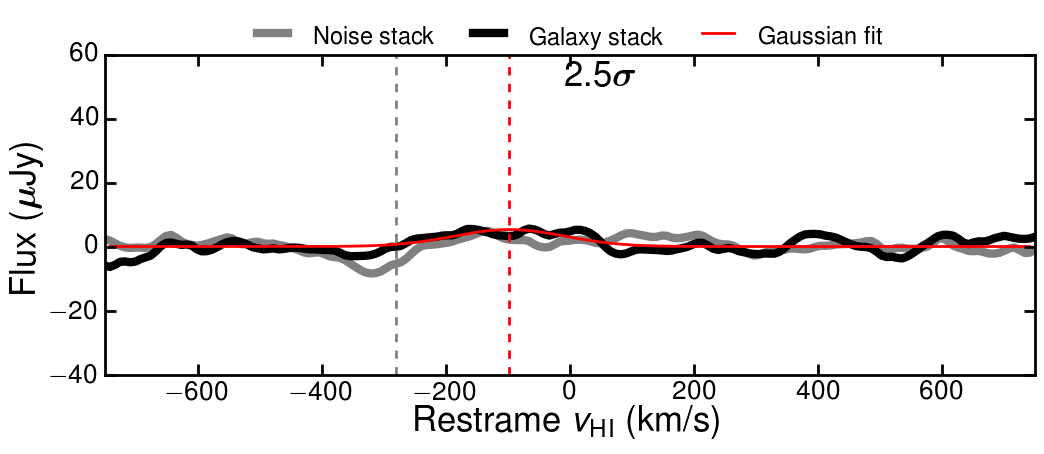}
\vspace{-10pt}
\caption{Cluster passive galaxies.}
\label{fig:stacks:passive}
\end{subfigure}
\begin{subfigure}[b]{0.495\textwidth}
\includegraphics[trim=0cm 0cm 0cm 0cm, width=0.995\textwidth ]{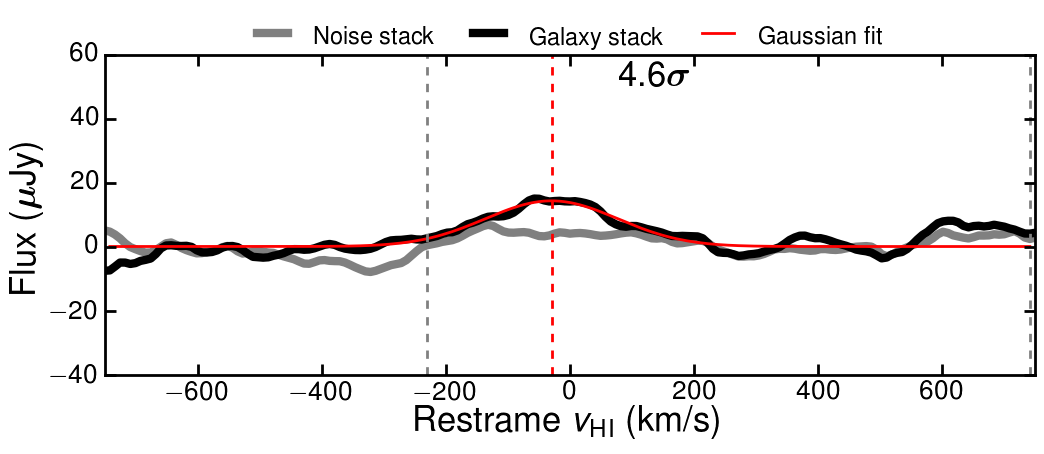}
\vspace{-10pt}
\caption{All galaxies.}
\label{fig:stacks:all}
\end{subfigure}
\caption{Final H{\textsc I} stacks after filtering and correcting for continuum over-subtraction. The solid black lines show the galaxy stacks and the gray lines show the corresponding noise properties. The thin, solid red lines show the Gaussian fits to the profiles located around a 0 restframe velocity. The vertical, dashed red line show the position of the peak, which is labelled with its significance based on the underlying noise properties. The RMS noise is calculated using only the data between the vertical, dashed gray lines, where at least $80\%$ of the peak number of sources are stacked. Note clear detections of H{\textsc I} are made for the line emitters, while no detections are made in the case of the cluster passive galaxies.}
\label{fig:stacks}
\end{figure*}

After the stacked spectra are filtered and corrected for the continuum over-subtraction, we fit Gaussian profiles around the 0-velocity position. Results are shown in Figure \ref{fig:stacks}. The parameters of the Gaussian fits can be found in Table~\ref{tab:stacked}. We calculate the RMS from the noise stack. Note that even though the number of galaxies in the cluster and field stacks is similar, the noise levels achieved are a factor of $\sim4$ different. This is because field galaxies are preferentially located at large distances away from the FOV centre, meaning that the noise levels at their location are higher. 

In the case of the H$\alpha$ emission-line galaxies, a clear peak is found around the 0-velocity position and a Gaussian profile could clearly be fit. In the cluster line emitter stack, we reach a noise level $\sigma_\mathrm{RMS}$ of $5.9$ $\mu$Jy and detect H{\textsc I} at a peak significance of $7.2\sigma$. The results are virtually unchanged if the 6 AGN are removed from the stack. No clear detection of H{\textsc I} is made in the case of the passive galaxies, despite the much larger number ($2.5$ times) of galaxies stacked, as compared to the line emitter sample (H{\textsc I} measured with a peak of $5.3$ $\mu$Jy, with a $\sigma_\mathrm{RMS}=2.1$ $\mu$Jy). The velocity position of the putative peak ($\sim-100$ km s$^{-1}$), highly offset from $0$, is likely a spurious peak and also indicative of a non-detection of H{\textsc I}. The high offset is highly unlikely to be caused by stripping, given we are averaging across $150$ passive galaxies within the cluster, which have random motions in the cluster potential. In some of the stacks (e.g. emission cluster galaxy stack and the emission line field galaxy stack), there are additional peaks off-centred from the 0-velocity. In theory, additional peaks in cluster stack, for example, could be caused by tidally stripped tails, pointing in the same redshift direction, such that they add coherently when stacked. However, this is highly unlikely as the cluster galaxies are moving on a range of orbits within the cluster potential. Even if H\textsc{I} tails existed, they would have a variety of orientations. When stacking the galaxies, the tails would therefore not add coherently. Therefore, we believe these to peaks to be caused by noise variations and low-level systematics. Note that at higher restframe velocities, the number of sources for which data exists dwindles (as shown in Figure \ref{fig:nostacked}). For example, at $\pm600$ km s$^{-1}$, the number of sources with data in that velocity channel is already half that at velocity 0. That means that the noise at larger restframe velocities will be higher than the noise around the 0-velocity. Therefore the significance of off-center peaks is actually very low (at least a factor $\sqrt{2}$ lower than if located at 0-velocity).

The stack using all the galaxies reaches a very low noise level of $3.1$ $\mu$Jy. The H{\textsc I} signal peaks at a significance of $4.6\sigma_\mathrm{RMS}$, value mainly due to the contribution of SF and AGN galaxies.

\begin{table*}
\begin{center}
\caption{Peak number of sources in every stack created for the different galaxy samples, the RMS noise value obtained for each stack and the parameters of the Gaussian fit to the H{\textsc I} signal. H{\textsc I} emission is securely detected, at its peak, in the emission-line galaxy and all galaxy stacks, but not in the cluster passive galaxies. When integrating the H{\textsc I} signal to estimate an H{\textsc I} mass, we obtain clear detections for the line-emitters.}
\vspace{-5pt}
\begin{small}
\begin{tabular}{l c c c c c c c c c c}
\hline
\hline
Sample & Number & $\sigma_\mathrm{RMS}$ & H{\textsc I} peak  & Peak significance & H{\textsc I} velocity & H{\textsc I} width & $M_{\rm HI}$ &  $M_\star$ & $M_{\rm HI}/M_\star$ \\   
  &  & ($\mu$Jy) & ($\mu$Jy) & $\sigma_\mathrm{RMS}$ & (km s$^{-1}$) & (km s$^{-1}$) & ($10^9$ M$_\odot$)  & ($10^9$ M$_\odot$) &  \\ \hline
Emission line, field & \phantom{$0$}$22$ & $21.3$ & $60.4$ & $2.8$  & \phantom{$-0$}$1.0$ & \phantom{$0$}$69.3$ & $1.86\pm1.20$  & \phantom{$0$}$4.8\pm0.8$ & \phantom{$0$}$0.39\pm0.26$ \\
Emission line, cluster & \phantom{$0$}$45$ & \phantom{$0$}$5.9$ &  $42.7$  & $7.2$ & \phantom{$-0$}$4.6$ & $137.0$ & $2.50\pm0.62$ & \phantom{$0$}$7.4\pm0.5$ & \phantom{$0$}$0.34\pm0.09$ \\
All emission-line  & \phantom{$0$}$67$ & \phantom{$0$}$8.2$ & $45.1$ & $5.5$ & \phantom{$0$}$-0.7$ & $104.7$ & $2.00\pm0.67$ & & \\
Passive, cluster & $154$ & \phantom{$0$}$2.1$  &  \phantom{$0$}$5.3$ & $2.5$ & $-97.4$ & \phantom{$0$}$90.1$ & $0.21\pm0.15$ & $25.6\pm0.4$ & $<0.008\pm0.006$ \\
All & $221$ & \phantom{$0$}$3.1$  &  $14.3$ & $4.6$ & $-29.5$ & $107.0$ & $0.60\pm0.30$ &  & \\
\hline
\end{tabular}
\end{small}
\vspace{-10pt}
\label{tab:stacked}
\end{center}
\end{table*}

\subsection{H{\textsc I} masses}
We use the following relation to convert from radio flux $S$ into H{\textsc I} mass \citep[$M_\mathrm{HI}$;][]{1959BAN....14..335W,1962AJ.....67..437R}:
\begin{equation}
\label{eq:MHI}
\frac{M_\mathrm{HI}}{\mathrm{M}_\odot} = \frac{236}{1+z} \left(\frac{D_L}{\mathrm{Mpc}}\right)^2 \left(\frac{\int S_V dV}{\mathrm{mJy}\;\mathrm{km}\;\mathrm{s^{-1}}}\right),
\end{equation}
where $M_\odot$ is the mass of the Sun, $z=0.192$ is the mean redshift of the sample of galaxies, $D_L = 940.7$ Mpc is the luminosity distance at that redshift and $\int S_V dV$ is the average of the H{\textsc I} emission over a restframe velocity range. As mentioned in \S\ref{sec:results:stacking}, the velocity is defined as $V = c (\nu_\mathrm{HI}-\nu_\mathrm{observed})/\nu_\mathrm{HI}$.

For the stacks with H{\textsc I} detections (Figure~\ref{fig:stacks}), the H{\textsc I} mass is averaged over $2\sigma_\mathrm{Gauss}$ ($2$ times the Gaussian dispersion) on either side of the peak position. For the passive population, we use the range within $200$ km s$^{-1}$ from 0-velocity position. The error in the H{\textsc I} mass is calculated by propagating the RMS error through equation \ref{eq:MHI}, using the same velocity range used for the integration of the signal.

Note that even though we use the filtered stacks to calculate the H{\textsc I} masses, similar results would be obtained even if the original data is used. Although the quality of the spectra improve, the errors on the final H{\textsc I} mass do not heavily depend on the filtering. As shown in equation \ref{eq:MHI}, the mass is effectively an integral over the profile. Hence, the effects of SG filtering are reduced because averaging over a velocity range equivalent to smoothing down to a resolution equal to that velocity range. 

We find that the average H{\textsc I} mass for the emission-line cluster galaxies ($M_\mathrm{HI} = (2.50\pm0.62)\times 10^9 M_\odot$) is a factor of $1.3$ higher than the mass of cold neutral gas in their field counterparts ($M_\mathrm{HI} = (1.86\pm1.20)\times 10^9 M_\odot$). However the difference is not significant. Additionally, the cluster galaxies are on average $\sim1.5$ times more massive than their field counterparts (see also Table~\ref{tab:stackedHA}). The average stellar masses ($M_\star$) are calculated over the same galaxies stacked for the H{\textsc I} analysis and the error reported is the standard deviation of the sample. Therefore, when accounting for the differences in stellar masses between the cluster and the field galaxies, the cluster line emitters are consistent with being as gas rich as the field counterparts. The fraction of neutral atomic gas to stellar mass $M_{\rm HI}/M_\star$ is $0.39\pm0.26$ and $0.34\pm0.09$ for field and cluster galaxies, respectively. This is a surprising result, as cluster galaxies have been found to contain less H{\textsc I} than galaxies in the field \citep[e.g.][]{2001ApJ...548...97S}.

The H{\textsc I} masses of the directly detected sources (Table~\ref{tab:radio}) are in the same range as the values for the average emission-line stacks (Table~\ref{tab:stacked}). However, we do not directly detect any of the emission-line systems. This indicates that we are biased towards detecting face-on sources, while the galaxies used for stacking have random orientation (with the bulk being oblique to edge on). 

Within our noise limits, we do not detect any significant H{\textsc I} within the passive cluster galaxy population ($M_\mathrm{HI} = (0.21\pm0.15)\times 10^9 M_\odot$). The passive population has at least nine times less H{\textsc I} gas than field line emitters (although not statistically significant) and about $12$ times less than the cluster star-forming galaxies and AGNs ($2.8\sigma$ significance, where $\sigma$ is calculated as the errors on the passive galaxy mass and the cluster line emitter mass, added in quadrature). Note, however, that the velocity range used for the integration of the H{\textsc I} signal is larger for the passive cluster members than what was used for the line-emitters. The cluster emission-line systems have on average much lower masses compared to the passive galaxies. Therefore, the ratio of H{\textsc I} to stellar mass for the passive cluster galaxies is less than $0.008$, a factor of$\sim30$ times lower than the cluster line emitters. Taking into account the errors on the gas fractions for the two populations summed in quadrature, the difference between the gas fraction in passive and cluster active galaxies is significant at the $4\sigma$ level.

\subsection{H$\alpha$ - H{\textsc I} correlation}

We compare the amount of cold gas and ionised content in each galaxy stack by investigating their H$\alpha$ ($L_{\mathrm{H}\alpha}$) and H{\textsc I} ($L_{\mathrm{HI}}$) luminosities.

The H{\textsc I} luminosity is calculated from the peak flux. The H$\alpha$ luminosity is calculated from the H$\alpha$ flux estimated from the NB observations (see \S~\ref{sec:obs:optical}), after correcting for extinction by Galactic dust, as well as for $1$ mag for intrinsic dust attenuation \citep{2012MNRAS.420.1926S} within each galaxy. We also remove the contribution of the adjacent [N\textsc{II}] line from the line flux \citep[for details please see][]{2014MNRAS.438.1377S,Stroe2015}. We average the corrected H$\alpha$ luminosities for galaxies within each stack to obtain a mean value for the ionised gas content as function of galaxy type. 

The luminosities are calculated in the following way:  
\begin{equation}
\label{eq:L}
L_{\mathrm{HI}, \mathrm{H}\alpha} = 4 \pi d_\mathrm{L} F_{\mathrm{HI}, \mathrm{H}\alpha} 
\end{equation}
where $F$ is the H$\alpha$ total flux and H{\textsc I} peak flux, respectively and $d_\mathrm{L}=940$ Mpc is the luminosity distance at the redshift of the cluster.

The H$\alpha$ can be converted into a SFR value using the relationship from \citet{1998ARA&A..36..189K}, which we correct for the \citet{2003PASP..115..763C} IMF, according to \citet{2007ApJS..173..267S}:
\begin{equation}
\label{eq:SFR}
\frac{SFR}{\mathrm{M}_\odot \mathrm{yr^{-1}}} = \frac{4.4 \times 10^{-42} L_{\mathrm{H}\alpha}}{\mathrm{erg\, s^{-1}}}.
\end{equation}

Note that not all H{\textsc I} stacked galaxies have an H$\alpha$ flux measurement (see Table~\ref{tab:Ha} for numbers). This may be because the sources have too faint H$\alpha$ line fluxes, below the limits of our NB H$\alpha$ survey. To test how the full H{\textsc I}  sample differs from the H{\textsc I} sample with H$\alpha$ measurements, we followed the procedure outlined in \S \ref{sec:results:stacking}-\ref{sec:results:HI} and stacked only the H{\textsc I} sources with H$\alpha$ measurements. We find that the peak H{\textsc I} fluxes and the average H{\textsc I} masses for subsamples with H$\alpha$ counterparts matches their parent sample within the error. Therefore, the subsamples with H$\alpha$ measurements are representative of the parent sample. Given the more robust measurement of the average H{\textsc I} properties for the full H{\textsc I} samples (driven by the higher number statistics), in comparing the H$\alpha$ and H{\textsc I} properties, we use average H{\textsc I} properties derived for the full samples. 

\begin{figure}
\centering
\includegraphics[trim=0cm 0cm 0cm 0cm, width=0.495\textwidth]{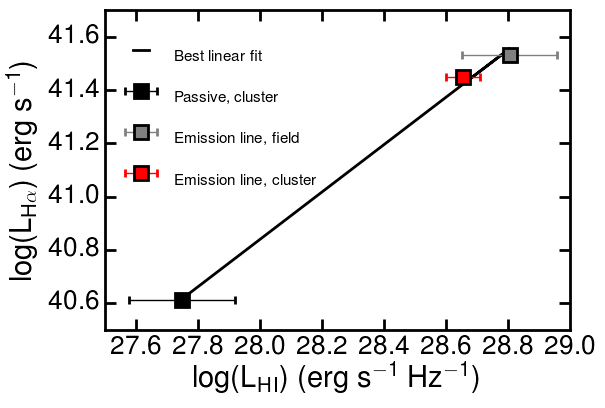}
\vspace{-10pt}
\caption{Relationship between H$\alpha$ luminosity and peak H{\textsc I} luminosity. As expected the cluster passive galaxies, selected to be non-line emitters contain less H{\textsc I} than emission-line galaxies. Line emitters are selected to be H$\alpha$ luminous, indicating the presence of vigorous SF and/or AGN activity.}
\vspace{-10pt}
\label{fig:HAvsHI}
\end{figure}

As Figure \ref{fig:HAvsHI} shows, the H$\alpha$ line emitters tend to be more luminous in H{\textsc I}. This is equivalent to galaxies which are more star-forming possessing larger reservoirs of atomic gas (Figure \ref{fig:SFRvsHI}). 
\begin{table}
\begin{center}
\caption{Optical properties of each galaxy stack. The average H$\alpha$ luminosities for each stack are calculated from data from \citet{Stroe2015}. The SFRs are calculated using the \citep{1998ARA&A..36..189K} conversion. Stellar masses are obtained using the method from \citet{Sobral2015}.}
\vspace{-5pt}
\begin{small}
\begin{tabular}{l c c c c}
\hline
\hline
Sample & Number & $\log_{10}(L_{\mathrm{H}\alpha})$ & SFR  & $M_\star$ \\   
  &  &  (erg s$^{-1}$) & ($M_\odot$ yr$^{-1}$) & ($10^9$ M$_\odot$) \\ \hline
Emission line, field & \phantom{$0$}$22$ & \phantom{$0$}$41.53$ & \phantom{$0$}$1.49$ &  \phantom{$0$}$4.8\pm0.8$ \\
Emission line, cluster & \phantom{$0$}$45$ & \phantom{$0$}$41.45$ & \phantom{$0$}$1.23$  & \phantom{$0$}$7.4\pm0.5$\\
Passive, cluster & $154$  & $<40.9$ & $<0.35$ & $25.6\pm0.5$ \\
\hline
\end{tabular}
\end{small}
\vspace{-10pt}
\label{tab:stackedHA}
\end{center}
\end{table}

\subsection{H$\alpha$ - radio correlation}

We extract sources from the VLA $1.5$ GHz image using {\tt PyBDSM} at the positions of the passive and H$\alpha$ line emitter galaxies with spectra (for number of sources see Table~\ref{tab:Ha}). The software detects single sources as islands and fits the flux distribution with Gaussians and also calculates the background noise levels using emission-free regions of the sky nearby each source. We assign a source a radio flux density by summing up the flux from all Gaussians belonging to its island. We cross-match radio sources with optical counterparts in our optical spectroscopic catalogue, using a maximum search radius of $5$ arcsec, to account for the positional accuracy of the optical and radio images as well as any extent the radio sources may have.

In case a source is not detected in the radio map, we assign it an upper limit flux equivalent to $3\sigma_\mathrm{RMS}$, where the $\sigma_\mathrm{RMS}$ is calculated from the noise level at the position of the source. Note that the FOV of the VLA image is large enough (FWHM of $\sim30$ arcmin) that is covers all the optical source positions. 

We calculate observed $1.4$ GHz measurements from the $1.5$ GHz values assuming a $-0.7$ radio spectral index value, and then convert the values to restframe $1.4$ GHz measurements. 
 
A plot showing the relationship between the H$\alpha$ luminosities and $1.5$ GHz luminosities (calculated using equation \ref{eq:L}) can be found in Figure \ref{fig:HAvsradio}. The fluxes of the radio counterparts and their morphologies can be found in the Appendix in Table~\ref{tab:radiocounter}. The emission-line galaxies have, on average, $1.5$ GHz luminosities $1-2$ orders of magnitude lower than the passive galaxies. Interestingly, even though the emission-line cluster members have similar H$\alpha$ luminosities, and hence SFRs, to the field line emitters, their radio BB detection rate is a factor of $>5$ higher. However, for all emission-line sources with radio detections, the H$\alpha$ luminosity correlates with the amount of radio emission. 

\section{Discussion}\label{sec:discussion}

\begin{figure}
\centering
\includegraphics[trim=0cm 0cm 0cm 0cm, width=0.495\textwidth]{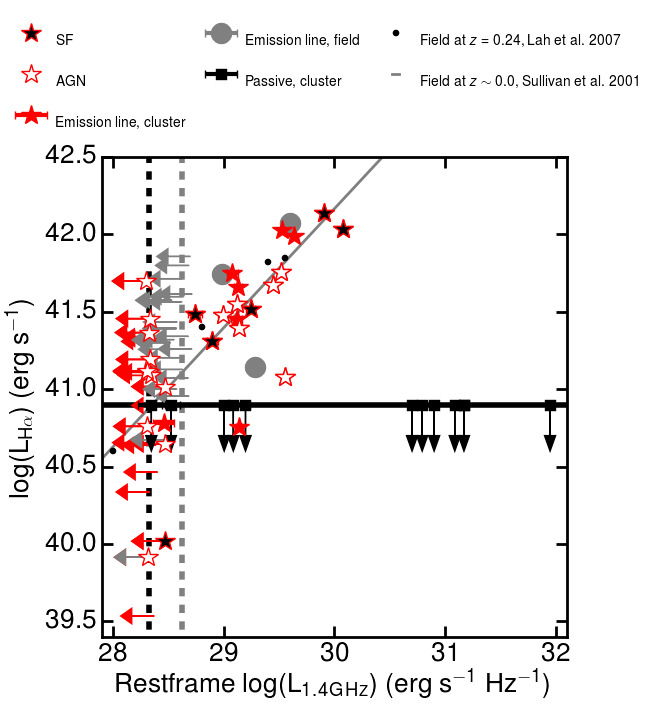}
\vspace{-10pt}
\caption{Relationship between H$\alpha$ luminosity and restframe $1.4$ GHz radio luminosity. The vertical black dashed line marks the RMS noise at the centre of the radio image, while the gray line represents the detection limit towards the edges of the FOV. The horizontal, solid, black line indicates the detection limit of the H$\alpha$ NB survey. Passive galaxies, undetected in H$\alpha$, are only plotted if they have a detection in the radio. Error bars are plotted but in most cases they are smaller than the symbol size. On average, the galaxies undergoing SF and optical AGN episodes have lower radio luminosities, probably driven by SNRs, while the passive cluster galaxies host powerful jetted and tailed radio sources. It is noteworthy that, in the case of line emission sources with radio detections the H$\alpha$ luminosity correlates well with the radio luminosity. We overplot for comparison data points for field galaxies at $z=0.24$ from \citet{2007MNRAS.376.1357L} and the linear correlation between H$\alpha$ and radio luminosity for local galaxies found by \citet{2001ApJ...558...72S}.}
\vspace{-10pt}
\label{fig:HAvsradio}
\end{figure}

Despite being extremely massive \citep[$M_{200}\sim 2\times10^{15} M_{\odot}$,][]{2015ApJ...802...46J, Dawson2015} and hot \citep[$T = 6-12$ keV,][]{2013MNRAS.429.2617O}, the `Sausage' merging cluster hosts numerous massive, H$\alpha$ emission-line galaxies displaying elevated levels of SF and AGN activity, outflows, high metallicities and low electron densities compared to galaxies in the field \citep{Sobral2015,Stroe2015}. In the present study, we find that the emission-line cluster galaxies have similar H{\textsc I} neutral gas fraction as the field galaxies. The data reveal linear correlations of the H{\textsc I}, H$\alpha$ line emission and radio BB luminosity (Figures \ref{fig:HAvsHI}, \ref{fig:HAvsradio}, \ref{fig:SFRvsHI}). By combining tracers of SF on different time scales, H{\textsc I}, H$\alpha$ and radio BB data, we can understand the circumstances under which the elevated activity can be triggered and also the possible future evolution of the SF properties in the cluster galaxies. 

\subsection{H{\textsc I} \& H$\alpha$ - tracing the gas that fuels future star formation episodes}
We make a clear detection of H{\textsc I} for the emission-line cluster galaxies giving an average mass of $(2.50\pm0.62) \times 10^9$ M$_\odot$, while the average H{\textsc I} mass for the field galaxies is $(1.86\pm1.20) \times 10^9$ M$_\odot$. \citet{Stroe2015} and \citet{Sobral2015} find that the stellar masses of H$\alpha$ cluster galaxies are on average higher than their field counterparts. For the samples used in the H{\textsc I} stacks, the cluster galaxies are about $1.5$ times more massive than the field line emitters (see Table~\ref{tab:Ha}).

Note that the cluster and field line emitters are selected in the same way and that the spectroscopic samples are representative of their parent samples \citep[see][]{Sobral2015}. Therefore the cluster and field emission-line galaxy samples within the `Sausage' field are fully comparable. 

As mentioned in \S~\ref{sec:intro}, previous studies of the H{\textsc I} content in cluster galaxies find that star-forming galaxies become increasingly H{\textsc I} deficient towards cluster cores, when controlling for stellar mass or optical disk size \citep[e.g.][]{2001ApJ...548...97S,2007ApJ...668L...9V, 2007MNRAS.376.1357L}. Contrary to previous work in the field, we find that our emission-line cluster galaxies are as gas rich as their field counterparts, despite the two samples having similar H$\alpha$ luminosities and hence similar SFR (see Table~\ref{tab:stackedHA} and Figure~\ref{fig:HAvsHI}). 

\begin{figure}
\centering
\includegraphics[trim=0cm 0cm 0cm 0cm, width=0.495\textwidth]{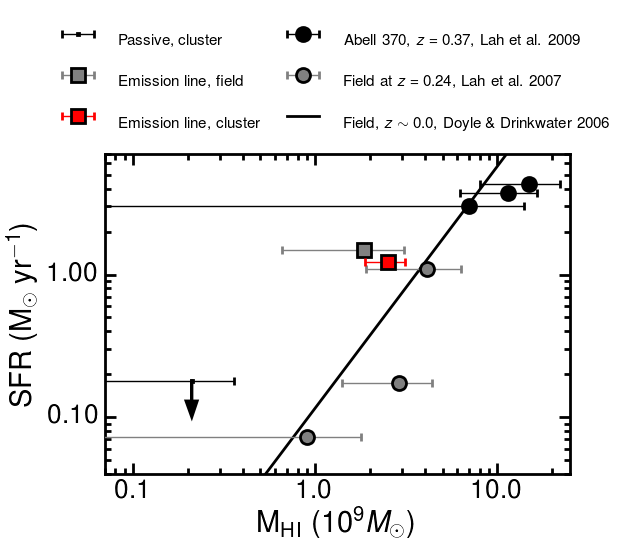}
\vspace{-10pt}
\caption{Relationship between SFR and integrated H{\textsc I} mass. All SFRs are calculated with a Chabrier IMF. For the passive population, the SFR is an upper limit. Overplotted are the data points for cluster Abell 370 at $z=0.37$ \citep{2009MNRAS.399.1447L} and a blank field at $z=0.24$ \citep{2007MNRAS.376.1357L}, together with the SFR-$M_{HI}$ relationship for the local Universe \citep{2006MNRAS.372..977D}. Note that it is difficult to compare our dataset with other work because of the different ways of measuring the H{\textsc I} mass and the different SF tracers used (H$\alpha$, [O{\textsc II}] or IR).}
\vspace{-10pt}
\label{fig:SFRvsHI}
\end{figure}

The studies of \citet{2009MNRAS.399.1447L} and \citet{2007MNRAS.376.1357L} indicate that a cluster at $z\sim0.37$ and a blank field at $z\sim0.24$ follow a similar relationship between SFR and H{\textsc I} mass to local field galaxies \citep{2006MNRAS.372..977D}. Even though our galaxies show evidence for a correlation between the amount of H$\alpha$ emission (or the SFR) and the H{\textsc I} mass, both the emission-line and passive galaxies do not follow the \citet{2006MNRAS.372..977D} relationship (see Figure \ref{fig:SFRvsHI}). The H{\textsc I} masses of our sample are $>5\sigma$ away from the masses predicted by the relationship at the same SFR. This could be entirely driven by the different SF tracers used in the different studies (\citet{2006MNRAS.372..977D} use infrared data, \citet{2009MNRAS.399.1447L} use [O\textsc{II}] and \citet{2007MNRAS.376.1357L} the H$\alpha$ emission line) or the spatial or velocity range over which the H{\textsc I} signal was integrated over.

Given the massive cluster galaxies may reside in massive dark matter haloes, they could have retained their cold H{\textsc I} gas more easily during the cluster merger. Interestingly, while spiral galaxies in the Virgo cluster are highly H{\textsc I} deficient, they are not deficient in molecular gas \citep{1986ApJ...301L..13K, 1989ApJ...344..171K, 1986ApJ...310..660S}. The authors attribute this to the preferential stripping of low-density gas located at the galaxy outskirts, therefore not affecting dense molecular gas located towards the galaxy centre. Therefore, in the case of the `Sausage' line emitters, with little to no ram pressure stripping of neutral and molecular gas, the larger reservoirs could fuel increased SFR in the cluster galaxies. If the cluster galaxies maintain their current average level of SF ($\sim1.23$ M$_\odot$ yr$^{-1}$, see Table \ref{tab:Ha}), and assuming $100\%$ efficiency in converting cold gas into stars, the H{\textsc I} reservoir would be depleted in $\sim2.0$ Gyr. If we assume a molecular gas content equal to the H{\textsc I} mass, it would take about $\sim4.0$ Gyr to consume the gas. However, as shown in \citet{Sobral2015}, the cluster galaxies also lose gas through outflows. Assuming a maximal mass outflow rate similar to the SFR \citep{2009ApJ...706.1364F, 2011ApJ...733..101G}, the H{\textsc I} gas will have been used up in about $\sim1.0$ Gyr ($\sim2.0$ Gyr if we include molecular gas). Assuming a more realistic outflow rate of about $0.1-0.5$ SFR (as observed by Swinbank et al. submitted), the H{\textsc I} gas would be depleted in $1.35-1.85$ Gyr, or $2.7-3.7$ Gyr if molecular gas is considered. This is in line with calculations from \citet{Stroe2015} where the molecular gas content was estimated using the total stellar mass, but atomic gas was not taken into account. 

\begin{table*}
\begin{center}
\caption{As Table~\ref{tab:stackedHA} but only for sources with H$\alpha$ measurements. Optical and broad-band radio properties of the sources in the line emitter and passive stacks. The average H$\alpha$ luminosities for each stack are calculated from data from \citet{Stroe2015}. The SFRs are calculated using the \citet{1998ARA&A..36..189K} conversion, with a \citet{2003PASP..115..763C} IMF.}
\vspace{-5pt}
\begin{tabular}{l c c c c c c}
\hline
\hline
Sample & Number H{\textsc I} sources with & $\log_{10}(L_{\mathrm{H}\alpha})^1$ & SFR$^1$ & Number H$\alpha$ & Number H$\alpha$ sources with & $\log_{10}(L_{\mathrm{H}\alpha})^2$ \\ 
  &  H$\alpha$ counterparts  & (erg s$^{-1}$) & $M_\odot$ yr$^{-1}$ & sources &  radio counterparts & (erg s$^{-1}$)  \\ \hline
Emission line, field & $20$ ($\sim91\%$) & \phantom{$0$}$41.53$ & \phantom{$0$}$1.49$ & $39$ & \phantom{$0$}$3$ (\phantom{$0$}$\sim8\%$) & $41.30$ \\
Emission line, cluster & $29$ ($\sim65\%$)  & \phantom{$0$}$41.45$ & \phantom{$0$}$1.23$ & $54$ & $24$ ($\sim44\%$) & $41.20$ \\
Passive, cluster &  $79$ ($\sim51\%$)  & $<40.9$ & $<0.35$ & $90$ & $11$ ($\sim12\%$) & $40.15$ \\
\hline
\end{tabular}\\
{\small $^1$ Average over the H{\textsc I} stacked sources with H$\alpha$ counterparts. $^2$ Average over all spectroscopic sources with H$\alpha$ measurements.}
\vspace{-10pt}
\label{tab:Ha}
\end{center}
\end{table*}

\subsection{Radio broad data - tracing the SN emission}
As shown by \citet{2001ApJ...558...72S}, synchrotron radio emission in star-forming galaxies is generated in super-nova remnants (SNR). Given the time required for a $7-8$ M$_\odot$ star to evolve to the red giant phase and undergo core-collapse, SNRs are good tracers of SF episodes happening $\sim100$ Myr ago \citep{1992ARA&A..30..575C}. 

In the case of H$\alpha$ luminous cluster galaxies, undergoing strong SF and optical AGN activity, the H$\alpha$ luminosity correlates well with the radio BB continuum (see Figure~\ref{fig:HAvsradio} and Table~\ref{tab:radiocounter} in the Appendix). The values fall on the same correlation as the large sample ($\sim350$) of typical $z=0.24$, H$\alpha$ luminous field galaxies studied by \citet{2007MNRAS.376.1357L} and follow the tight relationship found by \citet{2001ApJ...558...72S} for local field galaxies. For both samples, the galaxies classified as purely star-forming or as hosting an optical AGNs follow the H$\alpha$ - radio correlation. However, the AGN dominated cluster galaxies are expected to posses reasonable amounts of SF, as indicated by the spiral arm patters in many of their hosts \citep[for images see][]{Sobral2015}. Therefore, the sample includes photoionised broad line and narrow line regions \citep{Sobral2015} hosted by spiral galaxies, constituting examples of Seyfert galaxies. Even though some of the cluster galaxies are currently dominated by AGN, they have undergone significant SF activity in the past.

Despite their similar average H$\alpha$ luminosities, the fraction of H$\alpha$ luminous cluster members with radio counterparts is a factor $>5$ higher than their field counterparts, down to a similar sensitivity limit (see Table~\ref{tab:Ha} and Figure \ref{fig:HAvsradio}). This is consistent with results from \citet{Sobral2015} where they found evidence for strong outflows, probably driven by SN, only in the cluster galaxies and not in their field counterparts. Therefore, cluster galaxies have been undergoing SF for at least $100$ Myr, while field galaxies have been relatively inactive or less active in the past. Increased SF episodes tens to a few hundred Myr ago triggered by the cluster merger and its associated shock would also lead to higher SN rates. Field galaxies have not undergone any interaction with the shock front, hence in their case, there was no trigger for SN.

Overall, line emission galaxies both inside and outside the `Sausage' cluster follow the local relationship between H$\alpha$ and radio emission. Field emission line galaxies from a larger sample at $z\sim0.24$ \citep[$\sim150$ field galaxies][]{2007MNRAS.376.1357L} fall on the same relation. This indicates that the relationship between H$\alpha$ and BB radio emission does not evolve from $z\sim0.2$ to the present and that it does not depend on environment. The stellar populations in all galaxies that have been undergoing SF for longer periods ($\sim100$ Myr), irrespective of redshift or environment, seem to evolve similarly from the massive, short-lived stars which are responsible for producing the bulk of H$\alpha$ emission to the slightly less massive stars whose explosions dominate the SN population. These results indicate that the star formation history for the `Sausage' cluster galaxies is relatively constant, without any strong recent ($\sim10$ Myr) busts of star formation or in the past $100$ Myr.

A little bit over $10$ per cent of the passive galaxies have a radio BB counterpart, a similar rate to field emission line galaxies, much $4$ times lower than the cluster line emitters. The passive cluster members that have radio counterparts are giant ellipticals hosting radio jets and tails as indicated by the radio morphologies and luminosities. By contrast to the emission-line galaxies, where the radio emission is most likely produced by SNR, the radio emission in the elliptical galaxies traces shock-accelerated electrons in the jets and their back-flow (see also the Appendix). 

\subsection{Relationship to cluster merger state and shocks}

As the radio BB and H$\alpha$ data indicate, the `Sausage' cluster galaxies have been undergoing intense SF and AGN activity for at least $100$ Myr and this is likely to last a further $\sim1$ Gyr, given the large reservoirs of neutral hydrogen. The SF can last for an another $\sim2$ Gyr if comparable amounts of molecular gas are present. However, in the field galaxies around the cluster, we only find evidence of very recent SF episodes ($\sim10$ Myr). Despite the comparable amounts of H{\textsc I}, the cluster galaxies therefore underwent a significant event triggering SF about $\sim100$ Myr ago, evolving differently than their field counterparts. The most significant difference between the cluster and field line emitters is the cluster merger and the passage of the merger-induced shock waves only affected the cluster members.

The `Sausage' cluster is a result of a massive merger about $0.5$ Gyr ago \citep[e.g.][]{2011MNRAS.418..230V}, which produced shocks travelling through the ICM at about $2500$ km s$^{-1}$ \citep{2014MNRAS.445.1213S}. As the H{\textsc I} data indicate, the massive cluster members seem to have retained most of their neutral gas during the merger. Given their travelling speed, we expect the shocks to have traversed most cluster galaxies about $100-300$ Myr ago. The SF time scale imposed by the radio and optical tracers therefore matches well with the cluster merger time line. Our results fully support the interpretation previously proposed by \citet{Stroe2015} and \citet{Sobral2015}, where the cluster merger induced shocks trigger gas collapse as they traverse the gas rich cluster galaxies. This interpretation is also supported by simulations \citep{2014MNRAS.443L.114R}.  

\section{Conclusions}\label{sec:conclusion}
We presented deep H{\textsc I} observations combined with H$\alpha$ and broad band radio data to study the past, present and future SF activity in the `Sausage' merging cluster. Our main results are:
\begin{itemize}
\item The cluster H$\alpha$ emission-line galaxies (star-forming and radio-quiet broad and narrow line AGN), selected down to the same SFR limit, have as much H{\textsc I} gas ($(2.50\pm0.62)\times10^9\mathrm{M}_\odot$) as the field counterparts around the cluster ($(1.84\pm1.20)\times10^9\mathrm{M}_\odot$), when accounting for the different stellar masses of the two samples. This indicates the massive cluster line emitters retained their gas during the cluster merger.
\item A stringent upper limit is placed on the average H{\textsc I} content of the passive galaxies in the `Sausage' cluster: $\mathrm{M}_\mathrm{HI} = (0.21\pm0.15)\times10^9\mathrm{M}_\odot$. The ratio of H{\textsc I} to stellar mass for the passive galaxies is almost $40$ times less than for cluster line emitters (significant at $4\sigma$ level).
\item If the present SF and outflow rate is maintained in the emission-line cluster galaxies, their H{\textsc I} reservoirs will be depleted in $\sim0.75-1.0$ Gyr.
\item A large fraction of the emission-line cluster galaxies have radio BB detections, indicating the presence of SNR. These sources have been therefore undergoing vigorous SF for at least $100$ Myr.
\item The relationship between H$\alpha$ and radio continuum emission shows no evolution from $z\sim0.2$ to the present and also no dependence on environment.
\end{itemize}

Our H{\textsc I} observations represent an important milestone in the study of the `Sausage' cluster SF history. The member galaxies are gas-rich (gas to stellar mass ratio of $\sim0.34$) and thus capable of sustaining the increased SF and AGN activity measured in the cluster.

\section*{Acknowledgements}
We would like to thank the referee for their comments which greatly improved the clarity of the paper. We also thank Leah Morabito for useful discussions. This research has made use of the NASA/IPAC Extragalactic Database (NED) which is operated by the Jet Propulsion Laboratory, California Institute of Technology, under contract with the National Aeronautics and Space Administration. This research has made use of NASA's Astrophysics Data System. AS and HR acknowledge financial support from an NWO top subsidy (614.001.006). Part of this work performed under the auspices of the U.S. DOE by LLNL under Contract DE-AC52-07NA27344. DS acknowledges financial support from the Netherlands Organisation for Scientific research (NWO) through a Veni fellowship, from FCT through a FCT Investigator Starting Grant and Start-up Grant (IF/01154/2012/CP0189/CT0010) and from FCT grant PEst-OE/FIS/UI2751/2014.  RJvW is supported by NASA through the Einstein Postdoctoral grant number PF2-130104 awarded by the Chandra X-ray Center, which is operated by the Smithsonian Astrophysical Observatory for NASA under contract NAS8-03060. The Westerbork Synthesis Radio Telescope is operated by the ASTRON (Netherlands Institute for Radio Astronomy) with support from the Netherlands Foundation for Scientific Research (NWO). The Isaac Newton and William Herschel telescopes are operated on the island of La Palma by the Isaac Newton Group in the Spanish Observatorio del Roque de los Muchachos of the Instituto de Astrof\'isica de Canarias. Some of the data presented herein were obtained at the W.M. Keck Observatory, which is operated as a scientific partnership among the California Institute of Technology, the University of California and the National Aeronautics and Space Administration. The Observatory was made possible by the generous financial support of the W.M. Keck Foundation. Based in part on data collected at Subaru Telescope, which is operated by the National Astronomical Observatory of Japan. Based in part on observations from the Karl G. Jansky Very Large Array, operated by the National Radio Astronomy Observatory, a facility of the National Science Foundation operated under cooperative agreement by Associated Universities, Inc.

\bibliographystyle{mn2e.bst}
\bibliography{HI_sausage}

\appendix
\section{Radio fluxes and morphologies}
We tabulate the flux values of the 1.5 GHz VLA counterparts to the spectroscopic sources in Table~\ref{tab:radiocounter}. We also describe the morphology of the radio sources. Passive galaxies are hosts to jetted radio AGN, mainly pushed in wide or narrow angle tail morphologies by the interaction with the ICM. Emission line galaxies have mostly disk or spiral-like morphologies, indicating the radio emission is coming star formation.

\begin{table}
\begin{center}
\caption{Table with the radio fluxes, errors and morphologies of the VLA 1.5 GHz counterparts to the spectroscopic sources. The same sources tabulated here are plotted in Figure \ref{fig:HAvsradio}.}
\vspace{-5pt}
\begin{tabular}{c c c c c}
\hline
\hline
RA & DEC & Flux & Error & Morphology \\
(deg) & (deg) & (mJy) & (mJy) \\ \hline
 \multicolumn{5}{c}{Passive, cluster} \\						
340.6978	&	53.0939	&	15.034	&	0.022	&	NAT	\\
340.7687	&	53.1234	&	0.023	&	0.006	&	unresolved	\\
340.6048	&	52.9719	&	6.304	&	0.033	&	WAT	\\
340.7073	&	53.0401	&	0.034	&	0.007	&	unresolved	\\
340.8293	&	53.1228	&	12.420	&	0.094	&	WAT	\\
340.7072	&	53.0081	&	5.109	&	0.125	&	unresolved	\\
340.7051	&	53.0920	&	0.160	&	0.013	&	unresolved	\\
340.7191	&	53.0806	&	90.106	&	0.090	&	tailed	\\
340.7132	&	53.0139	&	8.084	&	0.053	&	WAT	\\
340.7385	&	53.1344	&	0.124	&	0.008	&	unresolved	\\
340.7720	&	52.9552	&	0.103	&	0.010	&	unresolved	\\ \hline
 \multicolumn{5}{c}{Emission line, field} \\	
340.3917	&	53.0163	&	0.194	&	0.014	&	unresolved	\\
340.6917	&	52.8561	&	0.404	&	0.025	&	unresolved	\\
340.9183	&	52.8727	&	0.097	&	0.013	&	unresolved	\\ \hline
 \multicolumn{5}{c}{Emission line, cluster} \\	
340.4575	&	53.0954	&	0.138	&	0.011	&	unresolved	\\
340.5930	&	53.0083	&	0.029	&	0.006	&	unresolved	\\
340.6151	&	52.9683	&	0.185	&	0.011	&	unresolved	\\
340.6518	&	53.0904	&	0.073	&	0.007	&	unresolved	\\
340.6598	&	53.0138	&	0.100	&	0.010	&	unresolved	\\
340.6642	&	52.9674	&	0.361	&	0.012	&	unresolved	\\
340.6711	&	52.9746	&	0.817	&	0.026	&	disk	\\
340.7136	&	52.9061	&	0.133	&	0.015	&	disk	\\
340.7225	&	52.9517	&	0.080	&	0.009	&	unresolved	\\
340.7369	&	52.9439	&	0.439	&	0.015	&	disk	\\
340.7401	&	53.0874	&	0.336	&	0.022	&	disk	\\
340.7475	&	53.0889	&	0.179	&	0.017	&	disk	\\
340.7502	&	53.0439	&	0.139	&	0.010	&	unresolved	\\
340.7796	&	52.9257	&	0.122	&	0.013	&	disk	\\
340.7870	&	53.0902	&	1.207	&	0.018	&	unresolved	\\
340.7941	&	53.0716	&	0.135	&	0.011	&	disk	\\
340.7985	&	53.0764	&	0.055	&	0.008	&	unresolved	\\
340.8037	&	53.0029	&	0.285	&	0.016	&	disk	\\
340.9204	&	53.0778	&	0.344	&	0.016	&	disk	\\
\hline
\end{tabular}\\
{\small NAT: narrow angle tailed galaxy; WAT: wide angle tailed galaxy; \\
Disk: galaxies with a disk or spiral-like morphology.}
\vspace{-10pt}
\label{tab:radiocounter}
\end{center}
\end{table}

\end{document}